%% file: paper.tex
\documentclass[sigconf]{acmart}

\usepackage{booktabs}
\usepackage{bbm}
\usepackage{makecell}
\usepackage[table]{xcolor} 

\begin{document}


\title{STCRank: Spatio-temporal Collaborative Ranking for Interactive Recommender System at Kuaishou E-shop}

\author{Boyang Xia}
\email{xiaboyang.tech@gmail.com}
\affiliation{%
  \institution{Kuaishou Technology}
\city{Beijing}
  \country{China}
}
\author{Ruilin Bao}
\email{baoruilin@kuaishou.com}
\affiliation{
\institution{Kuaishou Technology}
\city{Beijing}
  \country{China}
}

\author{Hanjun Jiang}
\email{jianghanjun@kuaishou.com}
\affiliation{%
  \institution{Kuaishou Technology}
\city{Beijing}
  \country{China}
}

\author{Jun Wang}
\email{wangjun03@kuaishou.com}
\affiliation{%
  \institution{Kuaishou Technology}
  \city{Beijing}
  \country{China}
  }



\author{Wenwu Ou}
\email{ouwenwu@gmail.com}
\affiliation{%
  \institution{Kuaishou Technology}
  \city{Beijing}
  \country{China}
  }


\renewcommand{\shortauthors}{Xia et al.}

\begin{abstract}
As a popular e-commerce platform, Kuaishou E-shop provides precise personalized product recommendations to tens of millions of users every day. To better respond real-time user feedback, we have deployed an interactive recommender system (IRS) alongside our core homepage recommender system. This IRS \footnotemark is triggered by user click on homepage, and generates a series of highly relevant recommendations based on the clicked item to meet focused browsing demands. 
Different from traditional e-commerce RecSys, the full-screen UI and immersive swiping down functionality present two distinct challenges for regular ranking system.
First, there exists explicit interference (overlap or conflicts) between ranking objectives, \textit{i.e.,} \textit{conversion, view} and \textit{swipe down}. This is because there are intrinsic behavioral co-occurrences under the premise of immersive browsing and swiping down functionality.
Second, the ranking system is prone to temporal greedy traps in sequential recommendation slot transitions, which is caused by full-screen UI design.
To alleviate these challenges, we propose a novel \textbf{S}patio-\textbf{T}emporal \textbf{C}ollaborative \textbf{Rank}ing (STCRank) framework to achieve collaboration between multi-objectives within one slot (spatial) and between multiple sequential recommondation slots. In multi-objective collaboration (MOC) module, we push Pareto frontier by mitigating the objective overlaps and conflicts. In multi-slot collaboration (MSC) module, we achieve global optima on overall sequential slots by dual-stage look-ahead ranking mechanism. Extensive experiments demonstrate our proposed method brings about purchase and DAU co-growth. The proposed system has been already deployed at Kuaishou E-shop since 2025.6.
\end{abstract}



\received{20 February 2007}
\received[revised]{12 March 2009}
\received[accepted]{5 June 2009}


\maketitle
\footnotetext{A demo video of Kuaishou Eshop IRS can be seen in https://youtube.com/shorts/iN8xbZ8db-I}
\input{intro}
\input{relatedwork}
\input{method}

\input{exp}
\newpage
\bibliographystyle{ACM-Reference-Format}
\bibliography{sample-base}
\end{document}

%% file: intro.tex
\section{Introduction}
\begin{figure*}
    \centering
    \includegraphics[width=0.9\linewidth]{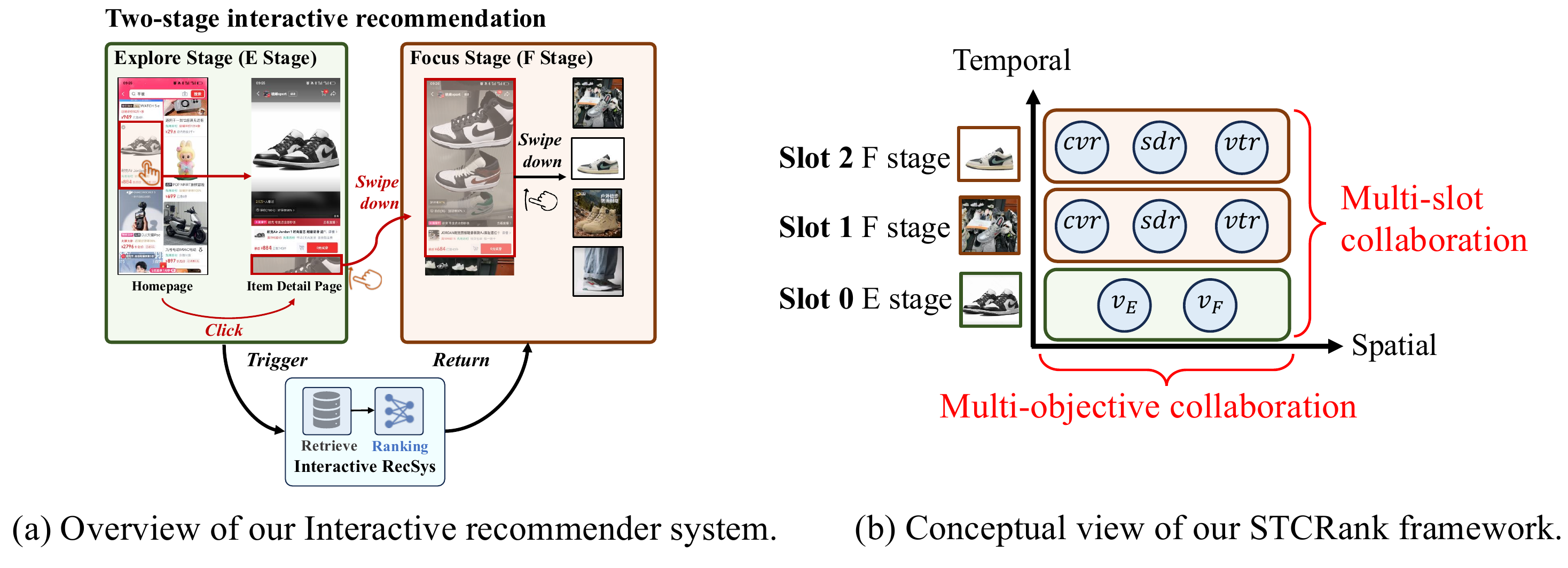}
    \caption{Overview of our interactive e-commerce recommender systems and STCRank framework.}
    \label{fig:main_fig}
\end{figure*}
Recommender systems (RS) effectively help people transform overwhelming data into personalized, contextually relevant knowledge streams.
However, conventional e-commerce recommendations struggle to capture and refine user interest signals (e.g., clicks) in real time. Interactive recommender systems (IRS) \cite{ali_irs,meituan_irs} solve this through a two-stage explore-focus mechanism: first probing broad user interests, then delivering targeted precommendations based on explicit feedback (e.g., the same category products). This approach bridges the gap between recommendation and search — offering more precision than traditional recommendations while remaining less effortful than search. Interactive recommendation has been widely adopted across domains, including: content feeds \cite{qnr,tutorial}, e-commerce \cite{dian,ali_irs}, local service \cite{meituan_irs}. 

IRS is particularly critical in e-commerce due to the inherent uncertainty associated with high-decision-cost non-standardized goods (e.g., clothing and electronics). Users often need repeated comparisons of brands, quality, and pricing to mitigate uncertainty before finalizing purchases \cite{santandreu2017purchase}. Consequently, users rarely convert on high-decision-cost items within a single recommendation slot. Thus, traditional e-commerce recommender systems struggle with high-involvement product conversions for high feedback latency. To address this, Kuaishou E-shop offers users a seamless transition from traditional recommendations to an immersive interactive recommendation flow (see Fig. \ref{fig:main_fig}(a)). 
To minimize user interruption, we intentionally abandon traditional IRS that captures user interests through explicit questioning UI \cite{qnr,tutorial}. Instead, we treat the entire homepage as an implicit `question list' (similar to \cite{meituan_irs}). The homepage recommendations are fully personalized based on user history, enabling users to shape their clear intention progressively through exploration (we call `Explore stage (\textbf{\textit{E stage}})' hereafter).
Upon clicking an item in E stage, an IRS is triggered and recommends relevant items based on immediate click signal after the normal item-detail page (see Fig. \ref{fig:main_fig}(a)). The IRS results are partially exposed at the bottom edge of the item-detail page, providing visual cues about the IRS without disrupting user's browsing the current item. The user only need to swipe down then they can actually enjoy highly focused recommendations (we call `Focus stage (\textbf{\textit{F stage}})'). Recommendation slots in F stage are exhibited in an immersive full-screen item detail UI, where users can finish conversion without explicit \textit{click and enter detail} procedure. 


A straight-forward approach to implementing the F-stage IRS is to deploy the traditional e-commerce recommender system of E stage.
In this way, we only need to constrain retrieval scope to relevant categories to the trigger item in \textit{retrieve module} and  align the specific ranking objectives in \textit{ranking module}.
However, this approach is suboptimal for our IRS due to two key challenges.
\textbf{(1) Objective interference caused by intrinsic behavioral correlations.} In conventional multi-objective RecSys, we expect no strong interference between objectives. Because strong interference, \emph{i.e.,} either overlap (positive) or conflict (negative) between objectives, make the ranking objective ensemble module either waste or suppress some objectives.
For an example of e-commerce RecSys, there is no strong interference between \textit{ctr} (click-through rate) and \textit{ctcvr} (post-exposure conversion rate), thus high \textit{ctr} does not imply high \textit{ctcvr} or low \textit{ctcvr}. 
 However in our F stage, there is a strong overlap between \textit{view} and \textit{conversion}, because long time viewing precede most purchase decisions. And there is a strong conflict between \textit{swipe down} and \textit{conversion}, because most users exit after conversion. These inferences necessitate a collaborative mechanism for multi-objective ranking module to advance the Pareto frontier. \textbf{(2) Temporal greedy trap induced by sequential recommendation slots.} The full-screen immersive UI of F stage forces strictly sequential transitions between recommendation slots. At the item detail page between E and F stages, if users do not swipe down, the pre-requested F-stage items will be wasted. Even upon entering F stage, only 1 of $m$ returned items is visible per view – exiting terminates subsequent exposures. This creates a greedy trap: when each slot greedily selects locally optimal items, holistic recommendation utility remain suboptimal. Thus it is critical to design a multi-slot collaborative mechanism to elevate the system’s overall ceiling.

To address these challenges, we propose \textbf{\textit{STCRank}}, a \textbf{S}patio-\textbf{T}emporal \textbf{C}ollaborative \textbf{Rank}ing framework that achieves holistic collaboration across both multiple objective within a slot \textit{spatially} and multiple sequential slots \textit{temporally} in the ranking system (see Fig. \ref{fig:main_fig}(b)).
From a spatial perspective, for each recommendation slot, our multi-objective collaboration (MOC) module resolves inherent overlaps and conflicts among \textit{view}, \textit{conversion} and \textit{swipe down} objectives, which effectively pushes the Pareto frontier.
From a temporal perspective, our multi-slot collaboration (MSC) module employs a look-ahead anticipation of subsequent exposure slots to achieve a global optima across slots, to mitigate the temporal greedy trap.
STCRank elevates the trade-off ceilings across both objectives and slots. Overall, our contributions can be summarized as:
\begin{itemize}
    \item We clearly characterize the unique challenges of the ranking system posed by special UI interaction mode in e-commerce IRS.
    \item We propose a novel \textbf{S}patio-\textbf{T}emporal \textbf{C}ollaborative \textbf{Rank}ing (STCRank) framework to mitigate interference among objectives and temporal greedy trap in multi-slot allocation.
    \item Extensive A/B experiments and empirical analyses demonstrate that our STCRank achieves co-growth of purchase and DAU. It can better satisfy users' demands on immersive browsing and buying in recommendations. STCRank has been fully deployed in Kuaishou's E-shop IRS since 2025.6.
\end{itemize}
 

%% file: relatedwork.tex
 \section{Related work}
 Interactive recommender systems (IRS) offer users an experience between passive recommendations and active search. It usually include 2 stages: firstly collecting users' intentions by question/topic list and secondly provide
precise recommendation based on collected feedback. 
\cite{qnr} propose a typical two-stage IRS to addresses the cold-start user recommendation problem. The first stage recommends potential topics of interest while the second stage delivers precise item recommendations based on the selected topics. \cite{ali_irs} argues that queries search ed by user historically serve as better indicators than manually defined topics since queries originate can reflect user preferences. Consequently, they replace the initial topic recommendation phase with query recommendations. \cite{meituan_irs} introduces an embedded interactive recommender system (EIRS) that transforms the entire initial homepage into a dynamic \textit{question list}. When users interact with the homepage, the EIRS is triggered to return a new item related to the interacted one. The new item will replace the subsequent originally recommended item in the homepage seamlessly. Our IRS synthesizes features from both traditional IRS (TIRS) and EIRS. Our first stage is similar to EIRS in \cite{meituan_irs}, which collects user intentions implicitly by utilizing the homepage as the \textit{question list}. However, our second stage is similar to TIRS, which provide precise recommendations through a new UI (full-screen immersive browsing UI) to follow user interactions. This design has two advantages. First, it eliminates the need for crafted question/topic recommendation in TIRS. Second, it surpass the limitations of EIRS that typically display only one IRS result - our solution supports continuous recommendation slots with immersive browsing experience.

%% file: method.tex
\section{Methodoloy}
\begin{figure*}
    \centering
    \includegraphics[width=0.9\linewidth]{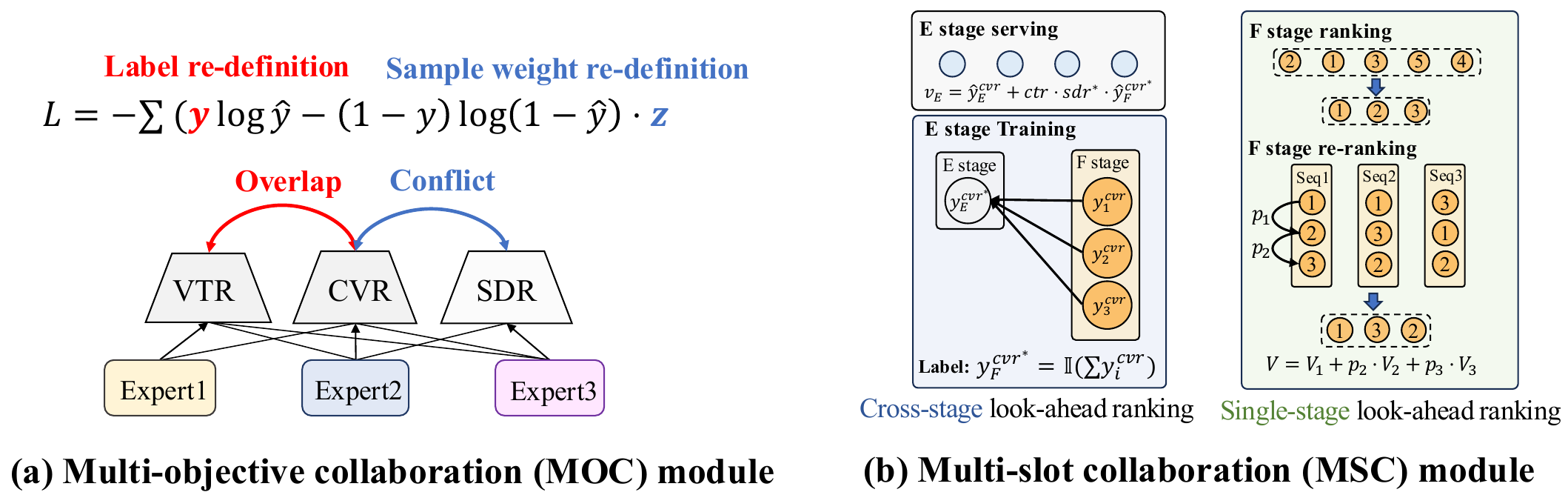}
    \caption{Overview of MOC module and MSC module.}
    \label{fig:main_fig_2}
\end{figure*}
\subsection{Problem formulation}
Before the ranking module, to provide users with candidates related to the trigger item (the clicked item in E stage), the \textit{retrieve} module retrieves $n$ candidates through multiple retrieve channels, \emph{e.g.,} trigger-based item2item retrieve and trigger category-constrained embedding-based user2item retrieve, \emph{etc}. 
The ranking module then selects the candidates with optimal comprehensive value to maximize dual objectives: purchase and DAU (in addition to DAU, we also utilize valid item-detail page views, \emph{i.e.,} IPV, to measure the strength change of user engagement induced by the strategies). A MMOE (Multi-gate Mixture of Experts) \cite{mmoe} based ranking model is trained to predict 3 objectives: conversion rate (\textit{cvr}), view-through rate (\textit{vtr}) and swipe-down rate (\textit{sdr}).
Here, the \textit{cvr} here aligns with traditional e-commerce’s post-exposure conversion rate (\textit{ctr} $\times$ \textit{cvr}). There is no concept of \textit{ctr} in F stage like in E stage because users are allowed to direct press the `buy' button without any `click \& enter in detail'  procedure (see Fig. \ref{fig:main_fig}). All objectives are optimized with binary cross entropy losses:
\begin{align}
\mathcal{L}(\Theta)=- \sum_{i\in \mathcal{S}} \sum_{j=1}^3 z^{j}_{i}\left(y^{j}_{i} \log \hat{y}^{j}_{i}-\left(1-y^{j}_{i}\right) \log \left(1-\hat{y}^{j}_{i}\right)\right),
\end{align}
where $y^{j}_{i}$, $\hat{y}^{j}_{i}$, $w^{j}_{i}$ are the label, prediction and sample weight for the sample $i$ of objective $j$, individually.
To sort all candidates, we ensemble all objectives via a linear additive formula and select the top-$m$ candidates in a point-wise manner.
\begin{align}
v[j]=& w_1 \cdot vtr + w_2 \cdot cvr + w_3 \cdot sdr,
\end{align} where $w_j$ is the weight for objective $j$. Then we elaborate on spatio-temporal collaborative framework below. 

\subsection{STCRank framework}
STCRank is composed of two modules, Multi-Objective Collaboration (MOC) module and Multi-Slot Collaboration (MSC) module. In MOC, we propose a suit of refined label design for \textit{view} and \textit{swipe-down} objectives, to effectively alleviate their inherent conflicts with \textit{conversion} objective. In MSC, we implement a multi-stage look-ahead ranking mechanism combining both cross-stage (E + F stage) and single-stage (F stage only) strategies, to optimize towards the global optima across recommendation slots.

\subsection{Multi-objective collaboration (MOC)}
In addition to conversion rate (\textit{cvr}), the primary ranking factor an e-commerce RecSys, the ranking model in F stage predicts two user engagement related objectives, view-through rate (\textit{vtr}) and swipe-down rate (\textit{sdr}) to encourage highly interacted items for DAU optimization. However, these factors are naturally conflict with \textit{cvr}, we use several strategies to achieve pareto optimization among these objectives.
\subsubsection{Reduce the overlap between view and conversion.}
Binary classification modeling with threshold discretization of viewing time is a popular solution for modeling \textit{vtr} factors. 
Before defining the binary classification threshold for \textit{vtr}, we adopt an in-depth analysis on the correlation between viewing time $T$ and \textit{cvr}. Intuitively, the longer a user views on an item, the higher the likelihood he buys it. However, this correlation does not hold uniformly across the distribution. As shown in Tab. \ref{tab:nonuniform}, the correlation (spearman's $\rho$) between \textit{conversion} and \textit{view} gradually strengthens as $T$ increases. This increasing correlation indicates overlap between them increases with $T$.
This observation indicates that a too large threshold results in large overlap between \textit{vtr} and \textit{cvr} and undermines the necessity of this factor. And too small threshold indicates low-strength interest feedback. 
\begin{table}[t]
    \centering
    \resizebox{0.37\textwidth}{!}{
    \begin{tabular}{c|c}
    \toprule
        $T$ buckets (second) & Spearman's $\rho$ between $y^{cvr}$ and $T$  \\ \midrule
        0\ -\ 2 & 0.0007  \\ \midrule
        2\ -\ 5 & 0.0068  \\ \midrule
        5\ -\ 25 & 0.0271  \\ \midrule
        >\ 25 & 0.0822  \\
        \bottomrule
    \end{tabular}
    }
    \caption{Non-uniform correlations between \textit{conversion} and \textit{view} objectives.}
    \label{tab:nonuniform}
\end{table}
Based on this observation, we search an appropriate discretization threshold within a small range empirically. We final determine 5 second as the threshold to define \textit{vtr} binary labels, which has less overlap \textit{conversion} than 25-second baseline and less conflict with \textit{conversion} than 2-second scheme.
\begin{align}
\text{baseline \textit{vtr} label:}\ y^{vtr}=\mathbbm{1}(T>25),\\
\text{MOC \textit{vtr} label:}\ y^{vtr}=\mathbbm{1}(T>5),
\end{align}

\subsubsection{Reduce the conflicts between swipe down and conversion.}
 In addition to view, swiping down is another import positive feedback. 
 When we trained a model with slide-down predictions and added this predicted \textit{sdr} to ranking mechanism straightforwardly, we found that not only did it fail to improve IPV, but it also exerts a significant decline on purchases.

\begin{table}[t]
    \centering
    \resizebox{0.3\textwidth}{!}{
    \begin{tabular}{c|cl}
    \toprule
        Session exit probability& $y^{cvr}=0$&$y^{cvr}=1$\\ \midrule
        Immediate exit& 19.7\%&61.3\%\\ \midrule
        Exit within 3 PVs& 33.6\%&74.8\%\\ \midrule
        Exit within 5 PVs& 42.9\%&78.8\%\\ \bottomrule
    \end{tabular}
    }
    \caption{Conflicts between \textit{swipe down} and \textit{conversion} objectives.}
    \label{tab:conflict}
\end{table}
After detailed data analysis, we find two key problems.
\textbf{(1) Low discrimination.} In a user session with $N$ PVs before session exit, there exist $N-1$ positive samples versus only 1 negative sample (the final exit action). This results in severely diminished discriminative power for \textit{sdr} prediction.
\textbf{(2) Label conflicts.} The swipe-down objective demonstrates significant conflicts with conversion objective. As shown in Tab. \ref{tab:conflict}, under 61\% situations, user exit (do not swipe down) immediately after conversion. This makes many positive training samples of conversion become negative ones for swiping down objective. Consequently, the predicted swipe-down factor by the model trained with this labels inherently suppress conversion objective. 

To address problem (1), we perform sample selection by retaining only the first-position PV, as the swiping down positive labels on the first position are much sparser than on all samples. To address problem (2), we redefine the negative samples for swiping down objective. We define the exit samples without conversion as negative samples while filter the exit samples with purchases - neither classified as positive nor negative samples.
\begin{align}
\text{baseline sample weights:}\ z^{sdr}_i&=1, i\in \mathcal{S},\\
\text{MOC sample weights:}\ z^{sdr}_i&=
\begin{cases}
0 & \text{if } y^{sdr}_i = 0 \text{ and } y^{cvr}_i = 1,\\
1  & \text{else.}
\end{cases}
\end{align}

\subsubsection{Multi-objective optimal hyper-parameter tuning.} 
We integrate the three ranking factors (vtr, sdr, cvr) into an additive formula, and treat the coefficients of each factor as tunable hyperparameters. We define the AUC (Area Under the Curve) sum of all factors as the optimization target and utilize an off-the-shelf Bayes optimization \cite{bayesian} based hyper-parameter optimization method to search the comprehensive best hyper-parameters. 

\begin{table*}[t]
\centering
\resizebox{1.0\textwidth}{!}{
\begin{tabular}{ccc|ccc}
\Xhline{0.8pt}
\textbf{A}&  \textbf{B}& \textbf{Modifications}& \multicolumn{3}{c}{\textbf{A/B Improvements}} \\
\hline
\multicolumn{3}{l|} {\cellcolor{gray!10}\textbf{Multi-objective collaboration module (MOC)}} & \cellcolor{gray!10} IPV (F) & \cellcolor{gray!10} Purchase (F) & \cellcolor{gray!10}DAU (F) \\
\hline
 MOC-base & Single obj. & Add \textit{vtr} objective.& +6.96\%& +0.53\%&+0.27\%\\ 
MOC-1& MOC-base& Reduce the overlap between \textit{vtr} and \textit{cvr}.& +4.87\%& +1.90\%& +0.15\%\\ 
MOC-2& MOC-1& Add \textit{sdr} and optimize the conflicts with \textit{cvr}.& +3.60\%& +0.42\%& +0.12\%\\ 
MOC-3& MOC-2& Pareto hyperparameter optimization. & +1.21\%& +4.18\%& +0.12\%\\
\hline
\multicolumn{3}{l|} {\cellcolor{gray!10}\textbf{Multi-slot collaboration module (MSC)}} & \cellcolor{gray!10} IPV (F) & \cellcolor{gray!10} Purchase (F) & \cellcolor{gray!10}DAU (F) \\
\hline
MSC-1& No MSC (MOC-3)& Add multi-slot look-ahead ranking in F stage.& +1.94\%& +2.10\%& +0.41\%\\ 
MSC-2& MSC-1& Add multi-slot look-ahead ranking across E and F stage. & +0.60\%& +2.66\%& +0.65\%\\ 
\hline
\multicolumn{3}{l|} {\cellcolor{gray!10}\textbf{Overall STCRank}} & \cellcolor{gray!10} IPV (E+F) & \cellcolor{gray!10} Purchase (E+F) & \cellcolor{gray!10}DAU (E+F) \\
\hline
MSC-2 & MOC-base & STCRank whole framework.& +9.65\% & +1.55\%& +0.03\%\\ 
\Xhline{0.8pt}
\end{tabular}
}
\caption{A/B test results of our proposed optimization strategies. `MSC' represents `Multi-Slot Collaboration' module and `MOC' represents `Multi-Objective Collaboration' module. `E stage' represents `Explore stage' and `F stage' represents `Focus stage'.}
\label{tab:main}
\end{table*}

\begin{table}[t]
\centering
\resizebox{0.49\textwidth}{!}{
\begin{tabular}{ccc|cc}
\toprule
\multicolumn{3}{c|}{A/B differences} & \multicolumn{2}{c}{A/B Improvements}\\ \midrule
A&  B& Modifications& IPV & Purchase \\ 
\hline\hline
SDR-1& No SDR& Add sdr.& +0.51\%& -3.22\%\\ 
SDR-2& SDR-1& \makecell[c]{First-position \\sample selection.}& +2.90\%& -5.95\%\\ 
 SDR-3& SDR-2& \makecell[c]{Conflict negatives \\filtering.}& +3.60\%&+0.42\%\\
\bottomrule
\end{tabular}
}
\caption{Ablation studies of swipe-down rate (sdr) optimization strategies.}
\label{tab:sdr}
\end{table}
\begin{table}[t]
\centering
\resizebox{0.27\textwidth}{!}{
\begin{tabular}{lcc}
\toprule
 Thre. &\multicolumn{2}{c}{A/B Improvements}\\ \midrule
     &IPV & Purchase \\ 
\midrule
    VTR(>25s)&-& -\\ 
    VTR(>5s)&+4.87\%& +1.90\%\\ 
     VTR(>2s)&+8.99\%&-0.49\%\\
\bottomrule
\end{tabular}
}
\caption{Ablation studies of choices of thresholds of view-through rate (vtr) model labels.}
\label{tab:vtr}
\end{table}

\subsection{Multi-slot collaboration (MSC)}
Existing recommendation systems usually treat temporally exposed recommendation slots as independent tasks and greedily maximize immediate purchases only at each recommendation slot. However, this greedy recommendation paradigm overlook the global objective optimization within a whole user session \cite{rerank}. For example, in our interactive recommendation system, the F stage recommendations are triggered based on E-stage feedback and optimizing the F stage for immediate conversion value does not guarantee joint optimality across both E-stage and F-stage.
\subsubsection{Cross-stage look-ahead ranking}
E stage recommendations indirectly influence the user intentions in F stage. In high-involvement non-standard categories, users exhibit strong demand for intra-category browsing and comparison. For these categories, immersive recommendations at the F-stage can help users quickly identify interests and finalize decisions. 
Intuitively, by recommending more products of these high-involvement categories at the E-stage, we can potentially improve the holistic purchase across both E stage and F stage.
Systematically, we focus on E stage ranking mechanism. For candidates with equivalent immediate conversion value at the E-stage, we should assign higher scores to those with higher potential conversion value at F-stage.

To model the look-ahead objective at the E-stage, we build a look-ahead label for E stage samples (see $y_F^{cvr^*}$ in Fig. \ref{fig:main_fig_2}) by aggregating all F-stage conversions as labels for E-stage samples.
\begin{align}
    y_F^{cvr^*} = \mathbbm{1}(\sum y_i^{cvr} > 0)
\end{align}
Since users traverse two funnel steps from E-stage exposure to F-stage conversion (click \& swipe down), directly modeling look-ahead objective from E-stage exposure space would encounter severe label skew. To address this, we decompose the prediction task into three factors: (1) E-stage click-through rate (ctr), E-stage→F-stage swipe-down rate (sdr$^*$), and within-F-stage conversion rate (cvr$^*$). Then we replace original immediate value based ranking objective by look-ahead value based ranking objective.
\begin{align}
    \text{Immediate value Ranking:}&\ \mathcal{R}_E =\text{argTopK}_{c \in \mathcal{C}_E} \hat{y}^{cvr}_E[c],\\
    \text{look-ahead Objective:}&\  V_F[c] =  ctr[c] \cdot sdr^*[c] \cdot \hat{y}^{cvr^*}_F[c],\\
    \text{look-ahead value Ranking:}&\  \mathcal{R}_E =\text{argTopK}_{c \in \mathcal{C}_E} \hat{y}^{cvr}_E[c] + V_F[c],
\end{align}
where $\mathcal{R}_E$ represents the returned results in E stage. $V_E$ represents value score of the given E-stage candidate, \textit{i.e., } conversion rate at E stage ($\hat{y}^{cvr}_E$ in Fig. \ref{fig:main_fig_2}. \ref{fig:main_fig_2}). 
\subsubsection{Single-stage look-ahead ranking}
Due to immersive full-screen UI constraints, when user is viewing a item card, the subsequent recommended items are totally invisible. Thus traditional greedy point-wise ranking increases the system's susceptibility to greedy traps for it inherently favors immediate conversions maximization. However, immediate conversions signify quick exits without additional future conversion opportunities. In this way, greedy ranking overlook the accumulation of long-term during user future swiping down and fail to obtain global optimum.

To address the aforementioned issues, we introduced a two-stage re-ranking \cite{rerank, prm} framework to select the best permutation after the ranking stage. The framework can be divided into two steps: sequence generation and evaluation. We firstly generate all possible $A_n^m$ permutations from the primary ranking results, where $n$ is the number of candidate items per request and $m$ is number of returned items to user interface. Then we compute the cumulative sequence value for each permutation and select the permutation with highest sequence value to return.

\textbf{Sequence evaluation.} 
The value of a given sequence $V[j]$ can be computed by the sum of item values $v[i]$ discounted by their exposure probabilities $p[i]$. The item value $v[i]$ here is defined as the same as  the ranking formula in ranking stage.
\begin{align}
V[j]=&\sum_i^m p[i]* v[i] \\
v[i]=& w_1 \cdot vtr + w_2 \cdot cvr + w_3 \cdot sdr
\end{align}
However, the position’s exposure probability is inherently dependent on all preceding items, which makes direct estimation challenging. To solve this, we decompose an exposure into a cumprod of condition exposure probabilties $P(\text{expo}_{i} \mid \text{expo}_{i-1})$, \emph{i.e.}, continuous swiping down rates. In this way, we transform exposure probability prediction into tractable per-slot \textit{sdr} prediction task, which we solved in the last section.
\begin{align}
p[i] & =P(\text{expo}_i \mid \text{expo}_{i-1})...  \cdot P(\text{expo}_{2} \mid \text{expo}_{1})\\
& =\prod_{k=1}^{i-1} sdr[k]
\end{align}

\textbf{Sequence generation.} 
To reduce computational complexity during the generation phase, we employ beam search \cite{beamsearch} for heuristic pruning. At each step of selecting subsequences of length $k \leq m$, only the top-\textit{B} highest-scoring subsequences are retained. The manner of value computation for  subsequences is as the same as the sequence evaluation phase.

%% file: exp.tex
\section{Experiments} 
\subsection{Metrics and details}
\textbf{Evaluation metrics.} We evaluate the effectiveness of strategies through three core metrics: IPV (valid Item-detail Page Views), purchase, and DAU (Daily Active Users).
The IPV here has stage-specific definitions. For the F stage, IPV refers to exposed PVs with viewing time exceeding 2 seconds. For the E stage, it measures actual clicked PVs. We utilize metrics of F-stage to measure individual strategies, \emph{e.g., } `IPV (F)'. We use combined metrics of E + F stage to measure the contribution to the holistic e-commerce feed of the platform, \textit{e.g.,} `IPV (F+E)'. All experiments share the same standardized testing protocols: each new experiment (A) uses the previous experiment's full-release version as baseline (B), with a fixed 7-day testing cycle.

\textbf{Implementation details.} The ranking models for E-stage and I-stage are jointly trained in an end-to-end manner. The user profiles, user history behavior, target item attributes (item ID, category, author) and trigger item (the item clicked in E stage) attributes are used as feature for the ranking models. The number of ranking candidates $n$ is 800 and the number of returned items $m$ is 10. The beam size $B$ we use in single-stage MSC is 25.

\subsection{A/B test results} 
Tab. \ref{tab:main} presents the related A/B improvements of different versions of the system  on IPV, purchases and DAU. 

\textbf{Results for MOC.} We first verify the impacts of original \textit{vtr} factor (trained with 25-second discretization labels), which improves IPV and DAU by a large margin (`MOC-base' v.s. `Single obj.'). 
Reduction of overlap between \textit{vtr} and \textit{cvr} objectives (`MOC-1') contributes 0.12\% DAU improvements over multi-objective baseline (`MOC-base'). Add \textit{sdr} and optimize the conflicts between \textit{sdr} and \textit{cvr} (`MOC-2') brings about additional 0.12\% DAU uplift. 
These experiments validates two key findings. (1) Accurately predicting user \textit{view} and \textit{swipe-down} behaviors, and incorporating these predictions into ranking objectives, can effectively enhance user interaction willingness with the system. (2) Reducing interference among objectives from the perspective of sample labels and weights (\emph{i.e.}, eliminating both overlap and conflict) can further elevate the Pareto frontier for multi-objective optimization.

\textbf{Results for MSC.}
As shown in Tab. \ref{tab:main}, single-stage MSC at F-stage (`MSC-1') presents 2.10\% purchase gain and a significant 0.41\% DAU uplift. The integrated cross-stage MSC between E and F stage (`MSC-2') further elevates  purchase by 2.66\% and DAU by 0.60\% respectively. 
The purchase gains demonstrate that our MSC mechanism elevates the upper bound of the whole ranking system. Meanwhile, MSC can better align with user expectations to browse more products before final decision-making at the F-stage than greedy ranking, which makes users more willing to enter from E-stage into F-stage

\subsection{Analysis}
\subsubsection{Choices of thresholds for binarizing \textit{vtr} labels.} 
As shown in Tab. \ref{tab:vtr}, through testing different thresholds for \textit{vtr} classification, we observed that lower threshold generates higher gains in IPV. However, lower threshold brings about a decline in purchases. This trade-off occurs because lower thresholds create a larger divergence from the purchase objective. In contrast, the 5-second view duration threshold strikes an optimal balance, simultaneously improving both purchases and IPVs.
\subsubsection{Ablation studies on sample weight strategies for \textit{sdr} model.}
The Tab. \ref{tab:vtr} demonstrates that directly incorporating the swipe-down rate factor not only fails to improve IPVs but also brings about a purchase decline (-3.22\%). When we apply first-position sample selection, we find marginal improvements on IPVs (+0.51\%) with larger negative effects on purchases. This verifies our hypothesis that this strategy enhances the discriminative power of the swipe-down rate. However, after filtering conflict samples, we successfully achieved IPV growth (+3.46\%) without purchase degradation. This demonstrates that the previous purchase decline results from the conflict negative samples.

\subsubsection{Analysis for cross-stage MSC}
As shown in Tab. \ref{tab:purchase}, we analyze the contributions on purchase gain of different categories. We find women clothes and women shoes accounted for the more percentages in total purchase increases than underwear and socks, which demonstrates the improvement in matching efficiency for high-involvement products is the key point of the effectiveness of our strategy.

\begin{table}[t]
\centering
\resizebox{0.3\textwidth}{!}{
\begin{tabular}{cc}
\toprule
Categories
& Contributions
\\ 
\bottomrule
Women's shoes
& +42.0\%\\ 
Women's clothes
& +25.2\%\\
Belts/hats/scarves
& +12.6\%\\
 Athletic shoes/clothing&+10.1\%\\
\bottomrule
\end{tabular}
}
\caption{Category breakdown of purchase gain.}
\label{tab:purchase}
\end{table}
\begin{table}[t]
\centering
\resizebox{0.3\textwidth}{!}{
\begin{tabular}{cc}
\toprule
Categories & Depth changes
\\ 
\bottomrule
All categories& +7.2\%\\ 
Women's shoes& +32.6\%\\
Bags and suitcases& +3.8\%\\
 Women's clothing&+2.6\%\\
 \bottomrule
\end{tabular}
}
\caption{Changes of session depth at conversion.}
\label{tab:depth}
\end{table}

\begin{table}[t]
\centering
\resizebox{0.25\textwidth}{!}{
\begin{tabular}{cc}
\toprule
TopK subsequence& HitRate
\\ 
\bottomrule
Hit@1& 30.9\%
\\ 
Hit@3& 62.1\%
\\
Hit@5& 73.8\%
\\
 Hit@All(10)&89.6\%\\
 \bottomrule
\end{tabular}
}
\caption{Hitrates@K between point-wise and permutation-wise ranking results.}
\label{tab:rerank}
\end{table}

\subsubsection{Analysis for single-stage MSC}
\begin{itemize}
    \item \textbf{Average session depth at conversion.} As shown in the Tab. \ref{tab:depth}, we observed an increase in users' average session depth at the purchase decision moment, particularly in women's shoes, bags and suitcases, and women's clothing categories. This indicates our method achieved an enhanced fulfillment of high-involvement purchase needs, where users demand more thorough evaluation before decisions.
    \item \textbf{Contributions of permutation changes \textit{v.s.} set changes.} 
Compared with original point-wise ranking, the re-ranking process introduced dual effects: (1) change of item set, \emph{i.e.,} the selected optimal set by re-ranking differ from the original point-wise ranking results. (2) permutation change, \emph{i.e.,} the internal ordering of the top items differ from original ranking results.
To evaluate the impacts of these two changes, we conducted topK subsequence consistency analysis using hitrate metrics. As shown in Tab. \ref{tab:rerank},  we find hitrate@All achieves almost 90\%, which indicates  the set-level divergence between re-ranking results and initial ranking results is negligible. However, hitrate@Top1 is only 31\%, which verifies that the gain almostly comes from the permutation changes.
\end{itemize}


\section{Conclusion}
We propose a STCRank framework for facilitate collaboration between different objectives within a slot (spatial) and different slots in temporal exposure sequence (temporal). Extensive experiments demonstrate that our proposed STCRank framework effectively resolves the multi-objective interference and temporal greedy trap problem, which improves both user engagement and conversion in Kuaishou's interactive recommender system (IRS). 